\documentclass{jfm}
\usepackage{graphicx}
\usepackage{import}
\usepackage{epstopdf, epsfig}
\usepackage{makecell}

\usepackage{xcolor}
\usepackage{verbatim}

\shorttitle{Flow Physics of COVID-19}
\shortauthor{R. Mittal, R. Ni and J-H Seo}

\title{The Flow Physics of COVID-19}

\author{
Rajat Mittal\aff{1,2},\corresp{\email{mittal@jhu.edu}},
 Rui Ni\aff{1} \corresp{\email{rui.ni@jhu.edu}}
 \and Jung-Hee Seo\aff{1} \corresp{\email{jhseo@jhu.edu}}
   }

\affiliation{
\aff{1}Department of Mechanical Engineering, Johns Hopkins University, Baltimore, MD 21218, USA
\aff{2}School of Medicine, Johns Hopkins University, Baltimore, MD 21218, USA }

\begin{document}

\maketitle

\begin{abstract}
Flow physics plays a key role in nearly every facet of the COVID-19 pandemic. This includes the generation and aerosolization of virus-laden respiratory droplets from a host, its airborne dispersion and deposition on surfaces, as well as the subsequent inhalation of these bioaerosols by unsuspecting recipients. Fluid dynamics is also key to preventative measures such as the use of face masks, hand-washing, ventilation of indoor environments, and even social distancing. This article summarizes what we know, and more importantly, what we need to learn about the science underlying these issues so that we are better prepared to tackle the next outbreak of COVID-19 or a similar disease.
\end{abstract}

\section{Introduction}\label{sec:Intro}

Transmission of respiratory infections such as COVID-19 is primarily via virus-laden fluid particles (i.e. droplets and aerosols) that are formed in the respiratory tract of an infected person and expelled from the mouth and nose during breathing, talking, coughing, and sneezing \citep{jones2015aerosol,social,asadi_coronavirus_2020,bourouiba2020turbulent}.
\citet{wells_air-borne_1934,wells_airborne_1955} showed that the competing effects of inertia, gravity, and evaporation determine the  fate of these droplets. Droplets larger than a critical size settle faster than they evaporate and contaminate surrounding surfaces. Droplets smaller than this size evaporate faster than they settle, forming droplet nuclei that can stay airborne for hours and may be transported over long distances.

Human-to-human transmission of COVID-19 occurs primarily via three routes: large droplets that are expelled with sufficient momentum so as to directly impact the recipients' mouth, nose or conjunctiva; physical contact with droplets deposited on a surface and subsequent transfer to the recipient's respiratory mucosa; and inhalation by the recipient of aerosolized droplet nuclei from the expiratory ejecta that are delivered by ambient air currents. The first two routes associated with large droplets are referred to as the ``droplet'' and ``contact'' routes of transmission, whereas the third is the so-called ``airborne'' transmission route \citep{jones2015aerosol}. Respiratory infections hijack our respiratory apparatus to increase the frequency and intensity of expiratory events such as coughing and sneezing, which are particularly effective in generating and dispersing virus-carrying droplets. 

Each stage in the transmission process is mediated by complex flow phenomena ranging from air-mucous interaction, liquid sheet fragmentation, turbulent jets, and droplet evaporation and deposition, to flow-induced particle dispersion and sedimentation. Thus, flow-physics is central to the transmission of COVID-19. Furthermore, 
given the centrality of flow phenomena to the transmission process, the methods, devices and practices employed to mitigate respiratory infections are also rooted in the principles of fluid dynamics. These include simple methods such as handwashing and wearing face masks, to fogging machines, ventilation \citep{tang2006factors} and even practices such as social distancing. However, despite the long history of medical research and experience in the transmission of respiratory infections (a fascinating account of the Spanish Flu can be found in \cite{soper2017lessons}), the rapid advance of COVID-19 around the world, has laid bare the limits of our knowledge regarding the physics underlying the transmission process, as well as the inadequacy of the methods/devices/practices used to curtail transmission rates. For instance, one factor that is contributing to the rapid growth of COVID-19 infections is the higher viral load of the SARS-CoV-2 virus in the upper respiratory tract of asymptomatic hosts who shed virus-laden droplets during normal activities such as talking and breathing \citep{bai2020presumed}. This knowledge gap has also manifested through guidelines on practices such as social distancing and the wearing of face masks \citep{NPR-facemask,Fortune_facemask}, which are based on outdated science \citep{asadi_coronavirus_2020,bourouiba2020turbulent,anchordoqui2020physicist}.

This article attempts to summarize our current state of knowledge, regarding the flow-physics implicated in the transmission of COVID-19. The challenge of summarizing such a vast topic is amplified by the need of the hour, that speed take precedence over comprehensiveness. Readers are therefore referred to other articles on this topic, \citep{asadi_coronavirus_2020,bourouiba2020turbulent,jones2015aerosol,johnson_mechanism_2009,tang_schlieren_2009,xie_how_2007,tang2006factors} to fill the many gaps that are sure to be left by this article.
\section{Respiratory Droplets and Aerosols}\label{sec:transport}
This section addresses the generation, expulsion, evolution, and transport of droplets and aerosols generated from the respiratory tract during expiratory activities such as: breathing, talking, coughing, and sneezing. The primary objective of fluid dynamic analyses in this setting is to  (a) determine the mechanisms for the generation of these droplets within the respiratory tract; (b) characterize the number density, size distribution, and velocity of ejected droplets; (c) determine the critical droplet size for transition between the large and small droplet transmission routes; (d) estimate the settling distance of large droplets; (e) determine evaporation times of small droplets; (f) characterize the transport of small droplets and droplet nuclei in the air; and (g) quantify the effect of external factors such as air-currents, temperature and humidity on all of the above. 

\subsection{Mechanisms of Droplet Formation }\label{sec:ejecta}
It is generally established that respiratory droplets are formed from the fluid lining of the respiratory tract \citep{morawska2005experimental,morawska_size_2009,johnson_mechanism_2009,edwards_inhaling_2004,almstrand_effect_2010}. Mechanisms of formation are usually associated with distinct locations in the respiratory tract; this is important because both the characteristics of the respiratory tract (length scales, airway elasticity, mucus and saliva properties etc.) as well as the  viral load carried by the lining are functions of the location \citep{johnson_modality_2011,almstrand_effect_2010}.

One key mechanism for the generation of respiratory droplets is the instability \citep{moriarty_flow-induced_1999} and eventual fragmentation of the mucus lining due to the shear stress induced by the airflow. Predicting the fragmentation and droplet size distribution resulting from this fragmentation is non-trivial because mucus is a viscoelastic shear-thinning fluid subject to surface tension. This enables multiple instabilities to bear on this problem \citep{malashenko2009propagation} including surface tension driven Rayleigh-Plateau instability \citep{plateau1873statique,rayleigh1878instability, rayleigh1879capillary,lin1998drop,eggers1997nonlinear}, shear driven Kelvin-Helmholtz instability \citep{thomson1871xlvi,von1868discontinuirliche,scardovelli1999direct,kataoka1983generation}, and acceleration-driven Rayleigh-Taylor instability \citep{lord1900investigation, taylor1950instability, joseph2002rayleigh,halpern2003nonlinear}. The Rayleigh-Taylor instability is particularly important in spasmodic events such as coughing and sneezing. 

The second mechanism for droplet formation is associated with the rupture of the fluid lining during the opening of a closed respiratory passage \citep{malashenko2009propagation}. One important site for this mechanism is in the terminal bronchioles during breathing. These sub-millimeter sized passages collapse during exhalation and the subsequent reopening during inhalation ruptures the mucus meniscus, resulting in the generation of micron-sized droplets \citep{almstrand_effect_2010,johnson_modality_2011}. A similar mechanism likely occurs in the larynx during activities such as talking and coughing, which involve the opening and closing of the vocal folds \citep{mittal2013fluid}. Finally, movement and contact of the tongue and lips, particularly during violent events such as sneezing, generates salivary droplets via this mechanism. The fluid dynamics of meniscus breakup associated with this mechanism are difficult to predict, especially given the non-Newtonian properties of the fluids involved, the dominant role of moving boundaries, and the large range of length- and time-scales involved in this phenomenon.

\subsection{Droplet Characteristics}
 The number density, velocity, and size distributions of droplets ejected by expiratory events have important implications for transmission, and numerous studies have attempted to measure these characteristics \citep{duguid_size_1946,wells_airborne_1955,han_characterizations_2013,morawska_size_2009,xie_exhaled_2009,bourouiba2014violent,scharfman_visualization_2016,asadi_aerosol_2019}. A single sneeze can generate $O(10^4)$ or more droplets with velocities upwards of 20 m/s \citep{han_characterizations_2013}. Coughing generates 10-100 times fewer droplets than sneezing with velocities of about 10 m/s, but even talking can generate about 50 particles per second \citep{asadi_aerosol_2019}. Measured droplet sizes range over four orders of magnitude: from $O(0.1)$ to $O(1000)$ microns. Recent studies have noted that while breathing generates droplets at a much lower rate, it likely accounts for more expired bioaerosols over the course of a day than intermittent events such as coughing and sneezing \citep{fiegel_airborne_2006,atkinson_quantifying_2008}. Consensus on all these droplet characteristics continues to be elusive due to the multifactorial nature of the phenomena as well as the difficulty of making such measurements \citep{han_characterizations_2013,chao_characterization_2009,morawska_size_2009}. 

\subsection{The Expiratory Jet and Droplet Transmission}
Droplets generated within the respiratory tract by the mechanisms described above are carried outward by the respiratory airflow, and those droplets that are not reabsorbed by the fluid lining are expelled within a two-phase buoyant jet from the mouth and/or nose. Breathing and talking generate jet velocities that seldom exceed 5 m/s \citep{tang2013airflow} and mostly expel small droplets. Violent expiratory events like coughing and sneezing on the other hand generate turbulent jets with Reynolds numbers of $O(10^4)$ and higher \citep{bourouiba2014violent}. Mucus and saliva that are expelled out of the nose and mouth can be stretched into ligaments and sheets and eventually fragmented into small droplets if the Weber number is large enough \citep{jain2015secondary}. This breakup process likely contributes to the generation of large droplets that fall ballistically and contaminate nearby surfaces \citep{bourouiba2014violent}.

 Wells' simple but elegant analysis predicted that the critical size that differentiates large from small droplets is about 100 microns \citep{wells_air-borne_1934}. Subsequent analysis has shown that typical temperature and humidity variations expand the critical size range from about 50 to 150 microns \citep{xie_how_2007}. For the droplet transmission route, an important consideration is the horizontal distance traveled by the large droplets. Thus the 3-6 ft. social distancing guidelines \citep{WHO-social-Distancing,cdc_social-distancing} likely originate from Wells' original work. 
 However, studies indicate that while this distance might be adequate for droplets expelled during breathing and coughing \citep{xie_how_2007,wei2015enhanced}, large droplets expelled from sneezes may travel 20 feet of more  \citep{xie_how_2007,bourouiba2014violent}. Studies also suggest that social distancing in indoor environments \citep{wong_cluster_2004} could be complicated by ventilation system-induced air-currents. 
 
 It has also been shown that the respiratory jet transforms into a turbulent cloud or puff \citep{bourouiba2014violent}. While large droplets are mostly not affected by the cloud dynamics, small and medium size droplets can be suspended in the turbulent cloud for a longer time by its circulatory flow, thereby extending the travel distance significantly \citep{bourouiba2014violent}. This also has important implications for transmission via indirect contact with contaminated surfaces and social distancing guidelines, since SARS-CoV-2 is able to survive on many types of surfaces for hours \citep{van_doremalen_aerosol_2020}. The turbulent cloud also moves upward due to buoyancy \citep{bourouiba2014violent}, thereby enabling small droplets and droplet nuclei to reach heights where they can enter the ventilation system and accelerate airborne transmissions (see Sec. \ref{sec:airborne}). The notion of a critical droplet size that was introduced by \cite{wells_air-borne_1934} might need to be reexamined in the light of our rapidly evolving knowledge about these expiratory events \citep{xie_how_2007,bourouiba2014violent}.

\subsection{Droplet Evaporation and Droplet Nuclei}\label{sec:evap}
Droplet evaporation plays a singularly important role in the eventual fate of a droplet \citep{wells_air-borne_1934}. The rate of evaporation depends on the difference between the droplet surface saturation vapor pressure and the vapor pressure of the surrounding air, the latter being dependent on humidity. The evaporation rate also depends on the mass-diffusion coefficient, which is a strong function of surface-to-ambient temperature difference, as well as the relative velocity between the droplet and surrounding gas. Thus, Reynolds, Nusselt, and Sherwood numbers for the droplets are just some of the non-dimensional parameters that determine this phenomenon \citep{xie_how_2007}. This dependence of evaporation rates on the ambient temperature and humidity has implications for the very important, and yet unresolved questions regarding seasonal and geographic variations in transmission \citep{tang2009effect,ma2020effects} rates as well as transmission in various indoor environments \citep{tang2006factors,li2007role}.

 Higher temperatures and lower relative humidities lead to larger evaporation rates that increase the critical droplet size \citep{wells_air-borne_1934,xie_how_2007}. However, temperature changes are usually accompanied by changes in humidity, and the overall effect of environmental conditions on transmission rates has been difficult to ascertain. This is not only due to the fact that these factors modulate the relative importance of the droplet and airborne routes of transmission, but also because survivability of enveloped viruses such as SARS-CoV-2 seem to be linked to these factors in a complex, non-monotonic manner  \citep{geller_human_2012}. Models that can combine droplet/aerosol fluid dynamics with virus microbiology and/or population dynamics could help unravel this complex effect of ambient conditions on transmission rates. 

\subsection{Airborne Transmission}\label{sec:airborne}
The airborne transmission route is associated with small droplets that are suspended and transported in air currents. Most of these droplets evaporate within a few seconds \citep{xie_how_2007} to form droplet nuclei, although the vapour-rich, buoyant turbulent expiratory jet can slow this evaporation \citep{bourouiba2020turbulent}. The nuclei consist of virions and solid residue \citep{vejerano_physico-chemical_2018} but water may never be completely removed \citep{mezhericher_theoretical_2010}. These droplet nuclei are submicron to about 10 microns in size, and may remain suspended in the air for hours. Each droplet nucleus could contain multiple virions, and given the approximately one hour viability half-life of the SARS-CoV-2 virus  \citep{van_doremalen_aerosol_2020} and the fact that SARS-type infections in a host may potentially be caused by a single virus \citep{nicas2005toward}, droplet nuclei play a singularly important role in the transmission of COVID-19 type infections \citep{asadi_coronavirus_2020}. 
The evaporation process of virus-laden respiratory droplets and the composition of droplet nuclei require further analysis because these have implications for the viability and potency of the virus that is transported by these nuclei.

The transport of  droplet nuclei over larger distances is primarily driven by ambient flows, and indoor environments such as homes, offices, malls, aircraft, and public transport vehicles pose a particular challenge for disease transmission. The importance of ventilation in controlling airborne transmission of infections is well known \citep{tang2006factors,li2007role} and much of the recent work in this area has exploited the power of computational fluid dynamic (CFD) modeling \citep{thatiparti_computational_2017,yang_effects_2018,yu_numerical_2018}. However, indoor spaces can have extremely complex flow domains, due not only to the presence of recirculatory flows driven by ventilation systems but also anthropogenic thermally-driven flow effects \citep{craven2006computational,licina2014experimental}. COVID-19 transmission from asymptomatic hosts \citep{bai2020presumed,ye2020delivery} makes it more critical than ever that we develop methods of analysis that provide better prediction of these effects. 

\section{Inhalation and Deposition of Bioaerosols}\label{sec:deposition}
The process of inhalation of virus-laden particles/droplets and deposition of the virus in the respiratory mucosa of the host is the final stage of airborne transmission. Fortunately, particle transport and deposition in the human airway has been studied extensively in the context of drug delivery \citep{heyder2004deposition}, food smell \citep{ni2015optimal}, and pollutant transport \citep{morawska2005experimental}.
The deposition of a solid particle is governed primarily by the mechanism of transport, whereas for liquid aerosols, the evaporation/diffusion process contributes significantly to the deposition mechanism. The latter is, however, a complex subject and has not been studied extensively so far \citep{rostami_computational_2009}. There are six mechanisms that determine the deposition location: impaction, sedimentation, interception, diffusion, electrostatic precipitation, and convection \citep{hinds1999aerosol}.
The relative importance of these mechanisms depends on the particle size and the region of the airway where deposition occurs. In general, for larger particles, impaction, sedimentation, and interception are more important than diffusion and convection\citep{rostami_computational_2009}.
For  droplet-nuclei-sized particles, sedimentation will drive significant deposition  in the upper respiratory tract of the host \citep{willeke1993aerosol}.  

Deposition of virus-bearing droplets on respiratory tract does not always result in infection, since the mucus layer provides some level of protection against virus invasion and subsequent infection \citep{zanin2016interaction}. The rate of droplet/nuclei deposition on the respiratory tract is quantified by the non-dimensionalized deposition velocity \citep{friedlander1957deposition, liu1974experimental}, which can vary by over four orders of magnitude \citep{guha2008transport}. For small droplets, deposition relies completely on turbulent diffusion \citep{friedlander1957deposition} but for large droplets, the deposition velocity increases substantially due to impact on the highly-curved and complex passage walls of the respiratory tract. Large droplets, despite a higher deposition velocity, likely deposit in the upper respiratory system, and could be deactivated by the first defensive layer of the mucosa \citep{fokkens2000upper}. On the other hand, droplet nuclei, despite their smaller deposition velocity, will penetrate deeper into the respiratory system, and this could affect the progression and intensity of the infection.

 Imaging modalities such as computed tomography (CT) and magnetic resonance imaging (MRI) provide realistic anatomical models for  experiments  \citep{ni2015optimal} and CFD models \citep{rostami_computational_2009}, from which local deposition can be quantified. A recent study even included the immune system response in the model \citep{haghnegahdar_lung_2019}, and similar models that combine fluid dynamics, biomechanics and virology, could serve as important tools in combating such pandemics.
 

\section{Measures to Mitigate Transmission}

\subsection{Mucus Property Modification}
Physical properties of the mucus play a key role in droplet formation within the respiratory tract. Transient modification of the physical properties of the mucus lining via material delivery to enhance mucus stability therefore provides a means for reducing infection rates. \cite{fiegel_airborne_2006} used isotonic saline to change the mucus lining properties via the induced ionic charge to reduce droplet formation, and \cite{edwards_inhaling_2004} explored the use of surface-tension-enhancing inhalants to reduce droplet generation. These techniques involve complex multiphysics flow phenomena that could benefit from advanced experimental and computational techniques.

\subsection{Fogging Machines}
Fogging machines provide an effective means for disinfecting large spaces, such as hospitals, nursing homes, grocery stores, and airplanes. Fogging machines that rely on the dispersion of a fine mist of disinfectants in the air have proven their performance in the healthcare sector \citep{otter2013role} and the food industry \citep{oh2005aerosolization}. Commercial fogging machines are also designed based on the same flow physics of aerosolization, and their droplet size is below 10 $\mu$m (Krishnan et al. 2012) in order to facilitate extended airborne duration.
 For this range of droplet sizes, it is likely that inertia-effects is small, and the collision between these disinfectant aerosol droplets and virus-bearing droplet nuclei might be dominated by diffusion. It is unclear if the collision rate will be enhanced by the turbulence generated by heating, ventilation, and air conditioning systems. Given that fogging machines have been widely employed in particle image velocimetry (PIV), experiments would be particularly well-poised to study these phenomena.

\subsection{Handwashing}
Transmission of infection from surfaces with virion-laden respiratory droplets usually occurs via hands \citep{nicas2009relative}, and handwashing with soap therefore remains the most effective strategy for mitigating this mode of transmission \citep{stock1940inactivation}. Soap molecules have a polar ionic hydrophilic side and a non-polar hydrophobic side that bonds with oils and lipids. Handwashing therefore works by emulsifying the lipid content of the material adhering to the hand into the bulk fluid and convecting it away. For enveloped viruses such as SARS-CoV-2, soap molecules also dismantle the lipid envelope of the virus thereby deactivating it \citep{kohn1980unsaturated}. The detritus from this disintegration is then trapped by the soap molecules into micelles, which are washed away. 

These molecular-level mechanisms are powered by macro-level flow phenomena associated with the movements of the hands. Amazingly, despite the 170+ year history of handwashing in medical hygiene \citep{rotter1997150}, we were unable to find a single published research article on the flow physics of hand-washing. The relative movement of the hands generates complex shear-driven flows of the soapy water, which forms a foam-laden, multiphase emulsion. Soap bubbles, which trap micelles, segregate rapidly from the fluid phase, thereby further accelerating the removal process. 

It is known that liquid foam exhibits elastic and plastic deformation under small and large stresses, respectively. With large enough deformation rates, the foam can rearrange its network and flow \citep{weaire1999physics}, which can be studied experimentally \citep{janiaud2005foam}. 
Reynolds numbers for these soapy liquid layers could exceed O(1000), suggesting that inertia, viscosity, surface tension, and gravitational forces would all play an important role in this process. An improved understanding of handwashing flow physics in the COVID-19 era could provide a science-based foundation for public guidelines/recommendations as well as new technologies that could improve the effectiveness of this practice.

\subsection{Face Masks}\label{sec:face_masks}
One issue that has generated significant controversy during the COVID-19 pandemic is the effectiveness of face masks \citep{Fortune_facemask,NPR-facemask,feng2020rational}. Indeed, it is likely that the years ahead will see the use of face masks become a norm in our lives. Understanding the physics that underpins the effectiveness of face masks as a defense against airborne pathogens is therefore, more important than ever.

Face masks provide ``inward'' protection by filtering virus-laden aerosolized particles that would otherwise be inhaled by an uninfected person; and ``outward'' protection by trapping virus-laden droplets expelled by an infected person \citep{van_der_sande_professional_2008}. The effectiveness of a simple face mask such as the surgical, N95 or homemade cloth face mask is a function of the combined effect of the filtering properties of the face mask material, the fit of the mask on the face, and the related leaks from the perimeter of the face mask. Each of these features implicates complex flow phenomena, which are briefly addressed here.

\subsubsection{Inward Protection}
The face mask material traps droplets and particles via the combined effects of diffusion, inertial impaction, interception, and electrostatic attraction \citep{thomas2016aerosol,fleming_keep_2020}. Filter efficiency (the ratio of the particle concentrations upstream and downstream of the mask) is a function of the particle and fiber size based Reynolds numbers, fiber-based Peclet number (for diffusion), particle-to-fiber size ratio (for interception), and Stokes number (for impaction). The  non-linear variation of  filtration mechanisms on these parameters generates a complex dependency of the filter efficiency on flow velocity, particle size, and filter material characteristics such as pore size, fiber diameter and electrostatic charge.

The process of inhalation generates a low pressure in the region interior to the face mask, thereby sealing (or at least reducing) perimeter leaks. Thus, with a reasonably well fit face mask, inward protection depends primarily on the face mask filter material. In this regard, an important characteristic of a face mask is the dependence of filtration efficiency on particle size.  Studies \citep{chen_aerosol_1992,weber_aerosol_1993,balazy_n95_2006} have shown that for a given filter, there is an intermediate particle size where filtration efficiency is minimum. Below this size, electrostatic attraction (active for masks such as N95) and diffusion dominate filtration, whereas above this size, impaction and interception are the dominant mechanisms. Aerosolized virus-laden droplets and droplet nuclei vary in size from sub-micron to millimeters and therefore the aforementioned size-dependent filtration efficiency is an important consideration for inward protection against COVID-19 infections. An increase in fiber density to enhance filtration efficiency is accompanied by an increased pressure drop across the mask \citep{lai_effectiveness_2012}, which requires a greater inhalation effort by the wearer. Thus, an appropriate balance has to be achieved via proper design of the filter material, and this might be particularly important in the post COVID-19 era, if wearing face masks becomes routine.

\subsubsection{Outward Protection}
The outward protection afforded by face masks has emerged as a particularly important issue in the COVID-19 pandemic because a SARS-CoV-2 
transmission may occur early in the course of infection, 
not only from symptomatic patients but also asymptomatic as well as minimally symptomatic patients \citep{zou2020sars,bai2020presumed,ye2020delivery}.
Indeed, the late switch to recommending universal use of face masks in the US \citep{NPR-facemask,Fortune_facemask} was based on the recognition that this spread by asymptomatic hosts might be a significant driver of COVID-19 infections. 

While a mask can significantly reduce the velocity of the throughflow jet during expiratory events \citep{tang_schlieren_2009}, the increased pressure in the region between the mask and the face pushes the face mask outwards, resulting in increased perimeter leakage \citep{liu1993respirator,lei_simulation_2013}. This fluid-structure interaction problem is mediated by the structural design as well as the permeability of the mask. The leakage jets that issue from the perimeter can be turbulent and highly directed (see for example, the flow visualization in  \cite{tang_schlieren_2009}), potentially serving as effective dispersers of respiratory aerosols in transverse directions. Spasmodic expiratory events such as coughing and sneezing that generate high transient expulsion velocities will significantly diminish the outward protection effectiveness of face masks \citep{lai_effectiveness_2012}. However, in a conceivable future where people will wear face masks while engaged in their daily routines, mask effectiveness during normal activities such as breathing and talking, might be equally important. 

In contrast to the problem of inward protection, which has been studied extensively \citep{chen_aerosol_1992,weber_aerosol_1993,balazy_n95_2006}, flow physics of outward protection from face masks is less well studied. \cite{tang_schlieren_2009} used Schlieren imaging to visualize cough-induced flow with and without face masks (surgical and N95). The study was extremely inventive but mostly qualitative, and future experiments should provide quantitative analysis of the leakage and throughflow jets, the aerosol dispersion through these jets, as well as the deformation of the mask during a variety of expiratory events. Recent CFD studies  of face mask aerodynamics \citep{lei_simulation_2013,hua_zhu_evaluation_2016} demonstrate the potential of computational modeling for this problem, but there is a critical need for modeling flow-induced billowing and associated leakage enhancement during expiratory events. Ultimately, analysis should not only enable a detailed evaluation of the protective efficiency of face masks; it should drive design changes that enhance mask performance and provide data that informs guidelines on practices such as social distancing.

\section{Closing}\label{sec:closing}
The COVID-19 pandemic has exposed significant scientific gaps in our understanding of critical issues ranging from transmission pathways of such respiratory diseases, to the strategies use for mitigating these transmissions. This article summarizes a fluid dynamicists' perspective on important aspects of the problem, including respiratory droplet formation, two-phase expiratory flows, droplet evaporation and transport, and face mask aerodynamics. COVID-19 touches almost every major arena of fluid dynamics, from hydrodynamic instability to porous-media and turbulent shear flows, from droplet breakup to particle deposition, and from Newtonian gas flows to non-Newtonian liquids. For the topics that we have discussed, breadth and medical context have taken precedence over a detailed exposition. COVID-19 has thrust the field of fluid dynamics into the public eye in a way \citep{parshina-kottas_this_2020,bourouiba2020turbulent} not seen since the space race of the 1960s. Our hope is that not only will this article serve as a call-to-arms to fluid dynamicists, it will also provide a starting point for the researcher who is motivated to tackle the science of COVID-19, and other similar diseases that are sure to appear in the not-too-distant future.

\section{Acknowledgements}
We gratefully acknowledge  Anjali Berdia, and Drs. Courtney McQueen, Andrea Prosperetti, H.S. Udaykumar, Gary Settles and Carolyn Machamer, who reviewed and commented on drafts of the manuscript.


\bibliographystyle{jfm}
\bibliography{References}

\end{document}